\documentclass[conference]{IEEEtran}
\IEEEoverridecommandlockouts
\usepackage{cite}
\usepackage{amsmath,amssymb,amsfonts}
\usepackage{algorithmic}
\usepackage{graphicx}
\usepackage{textcomp}
\usepackage{xcolor}
\def\BibTeX{{\rm B\kern-.05em{\sc i\kern-.025em b}\kern-.08em
    T\kern-.1667em\lower.7ex\hbox{E}\kern-.125emX}}
\begin{document}

\title{A Reference Architecture for Smart \\ and Software-defined Buildings}

\author{\IEEEauthorblockN{Manuel Mazzara}
\IEEEauthorblockA{\textit{Innopolis University, Russia} \\
m.mazzara@innopolis.ru}
\and
\IEEEauthorblockN{Ilya Afanasyev}
\IEEEauthorblockA{\textit{Innopolis University, Russia} \\
i.afanasyev@innopolis.ru}
\and
\IEEEauthorblockN{Smruti R. Sarangi}
\IEEEauthorblockA{\textit{IIT Delhi, India} \\
srsarangi@cse.iitd.ac.in}
\and
\IEEEauthorblockN{Salvatore Distefano}
\IEEEauthorblockA{\textit{University of Messina, Italy} \\
sdistefano@unime.it}
\and
\IEEEauthorblockN{Vivek Kumar}
\IEEEauthorblockA{\textit{NUST-MiSiS, Russian Federation} \\ 
vivekkumar0416@gmail.com}
}

\maketitle

\begin{abstract}
The vision encompassing Smart and Software-defined Buildings (SSDB) is becoming more and more popular and its implementation is now more accessible due to the widespread adoption of the IoT infrastructure. Some of the most important applications sustaining this vision are energy management, environmental comfort, safety and surveillance. This paper surveys IoT and SSB technologies and their cooperation towards the realization of Smart Spaces. We propose a four-layer reference architecture and we organize related concepts around it. This conceptual frame is useful to identify the current literature on the topic and to connect the dots into a coherent vision of the future of residential and commercial buildings.
\end{abstract}


\section{Introduction}
\label{sec:intro}
%\textbf{TODO: retarget in the smart buildings context.}
Smart and Software-defined Buildings (SSDB) represent the introduction of hardware, software and sensing into the places where we live, in the same way as electronics has been introduced into cars and vehicles over the last twenty years. It is therefore the new frontier for what concerns housing solutions \cite{SMForbes}. 
Advanced technology was introduced in cars in 80s initially in expensive models for then becoming the normality for small-size cars, and eventually legal frameworks adjusted in order to make formally mandatory the presence of certain sensing and acting functions \cite{ElectronicAutomotive}. In the same way, technology will possibly enter the market starting from high-end housing and public places to then move to more popular solutions. 
This paper surveys IoT and SSDB technologies and their cooperation towards the realization of Smart Cities and Smart Environments. 
We propose a four-layer reference architecture for SSDB and we organize related concepts around it. 
This conceptual frame is useful to identify the current literature on the topic and to connect the dots into a coherent vision of the future of residential and commercial buildings.

%\textbf{TODO: highlight contribution. 
%1 - overview of the topic, 2 - reference architecture, 3 - ???, ...}

The manuscript is organized as follows: after this introduction of the domain in Section \ref{sec:intro}, we first define Smart and Software Defined Buildings (Section \ref{sec:SDB}) and discuss the hardware infrastructure (Section \ref{sec:IoT}) and the communication networks  and protocols (Section \ref{sec:net}) that are necessary for their realization. 
We then overview the SSDB management in Section \ref{sec:mgmt}, their applications and services in Section \ref{sec:app} and crosscutting concerns in Section \ref{sec:cc}. 
Finally, in Section \ref{sec:conclusions} we present our conclusion.

\section{Smart and Software Defined Buildings}
\label{sec:SDB}
%\textbf{TODO:
%\begin{itemize}
%    \item INTRODUCE RELATED CONCEPTS such as: Smart Home, Smart Office, smart shops, Smart Spaces and their relationships with smart buildings (Venn diagram?) - DONE (MM)
%    \item Overlapping, Boundaries are blurred  $->$ Need of an orchestrated approach to (co-)design SH, SO, SS and SB. Could a smart building include some of them?
%    \item architecture of a smart building ICT Infrastructure: 4-5 layers
%    \begin{itemize}
%       \item Hardware - (IoT) Devices 
%      \item Network
%       \item Management of resources (interoperability, pooling, %scheduling, ..., eg OS OpenHab, BoS, BOSS, ...) and data (storing, %analytics, bigdata,)  building management system or building automation system (BAS).
%       \item applications and services
%       \item crosscutting concern: security, QoS, billing?, ...
%    \end{itemize}
%\end{itemize}
%}
%\textbf{TODO: add a definition of SB}
%Smart Buildings
Smart buildings (SB) are structures that use automated processes to control operations such as heating, ventilation, air conditioning, lighting, and security, and allowing sophisticated monitoring and control over their functions \cite{8418047}. One of the major set of functionalities that can be managed is the one concerning the \textit{environment and users' comfort}. This includes temperature, light, and humidity.  More complex functions can also be performed such as \textit{presence monitoring, identity recognition and detection of users' emotional states} \cite{Nalin2016}, 
%. This has 
under concomitant legal and ethical considerations.
SB also represent an important element in a Smart City ecosystem, and are therefore often considered as drivers for the \textit{urban smartization} process.  

%\textbf{TODO: add a definition of SDB}
%human-centered cyberphysical system making use of the IoT infrastructure combined with Artificial Intelligence and Machine Learning in order to maximize users' satisfaction and expectations. 
%Software-Defined Buildings 
A Software Defined Building (SDB)  \cite{SDB1} is a \textit{``programmable''} building abstracting and virtualizing  \cite{SDB2} the underlying (ICT) physical infrastructure to make it available (through specific API) for third party applications and services. 
SDB provide a \textit{Building Service Layer} to implement an interoperable and programmable environment for their management and to control applications and services, following a software defined approach also adopted in Smart Cities \cite{SDC}. 
Smart and Software Defined Building  are two related concepts, even if not strictly dependent on each other. 
An SB can be either based on an SDB or not, and an SDB does not necessarily imply an SB. 
In the former case of an SB established on top of an SDB,  specific SB services and applications for the building management are deployed on the SDB infrastructure by injecting their codes on the SDB programmable devices and nodes.
An SDB-powered SB inherits the benefits of programmability from SDB, allowing to reuse and share existing physical resources and infrastructure through different services and applications. 
This allows that existing applications and services can be easily evolved and/or maintained or even extended with new  ones, always exploiting the same SDB infrastructure.

\begin{figure}[htbp!]
    \centering
    \includegraphics[width=.6\columnwidth]{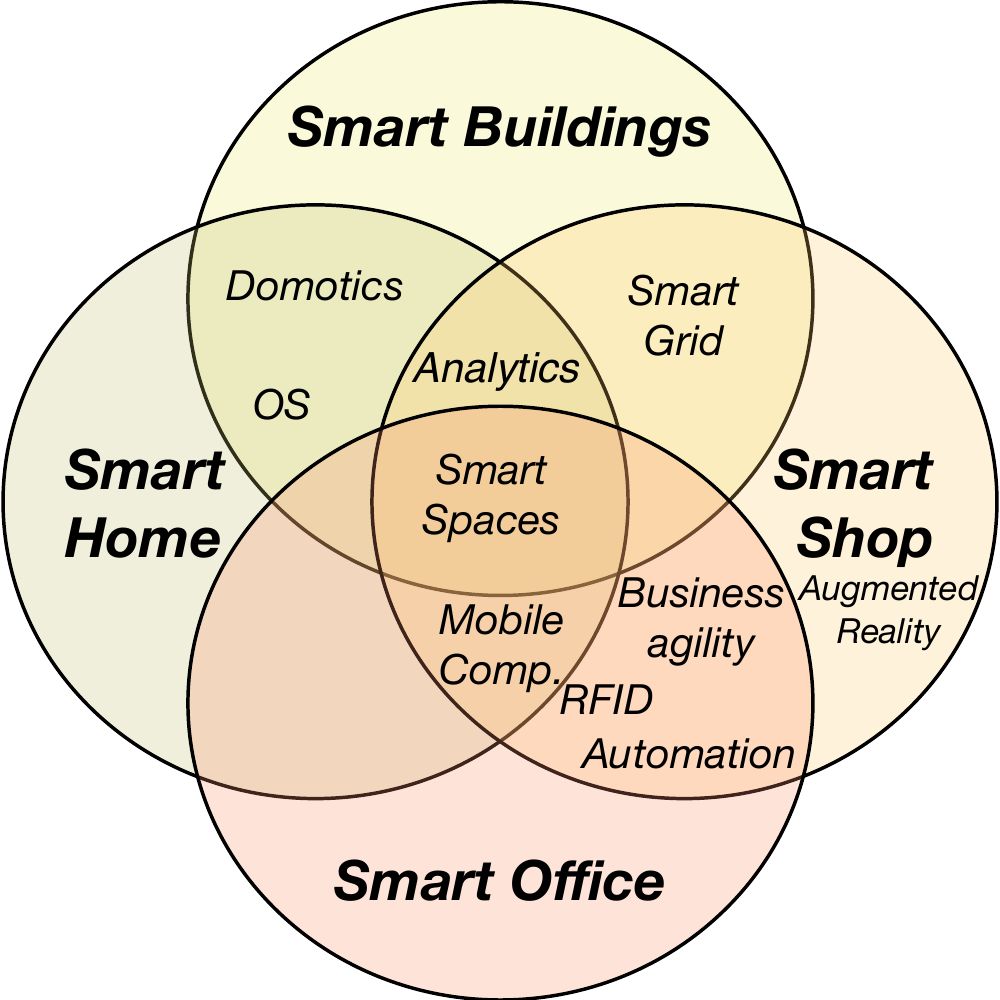}
    \caption{Relationships among Smart Home, Office, Shop, Building concepts.}
    \label{fig:sbhps}
\end{figure}

In the SSDB domain we can identify at least three different concepts that can be related to SB: \textit{smart home}, \textit{smart office} and \textit{smart shop}. Their interactions and overlapping are shown in Fig. \ref{fig:sbhps}. The overall idea lying behind all these concepts is the one of \textit{smart space}, i.e. a technology-equipped environment (residential, commercial or governmental) that facilities life and operations of the people who are living it. \textit{Domotics} and \textit{Automation} are essential parts of smart spaces, as well as data management and \textit{Analytics}.

\subsection*{SSDB Reference architecture}

\begin{figure}[ht]
    \centering
    \includegraphics[width=.7\columnwidth]{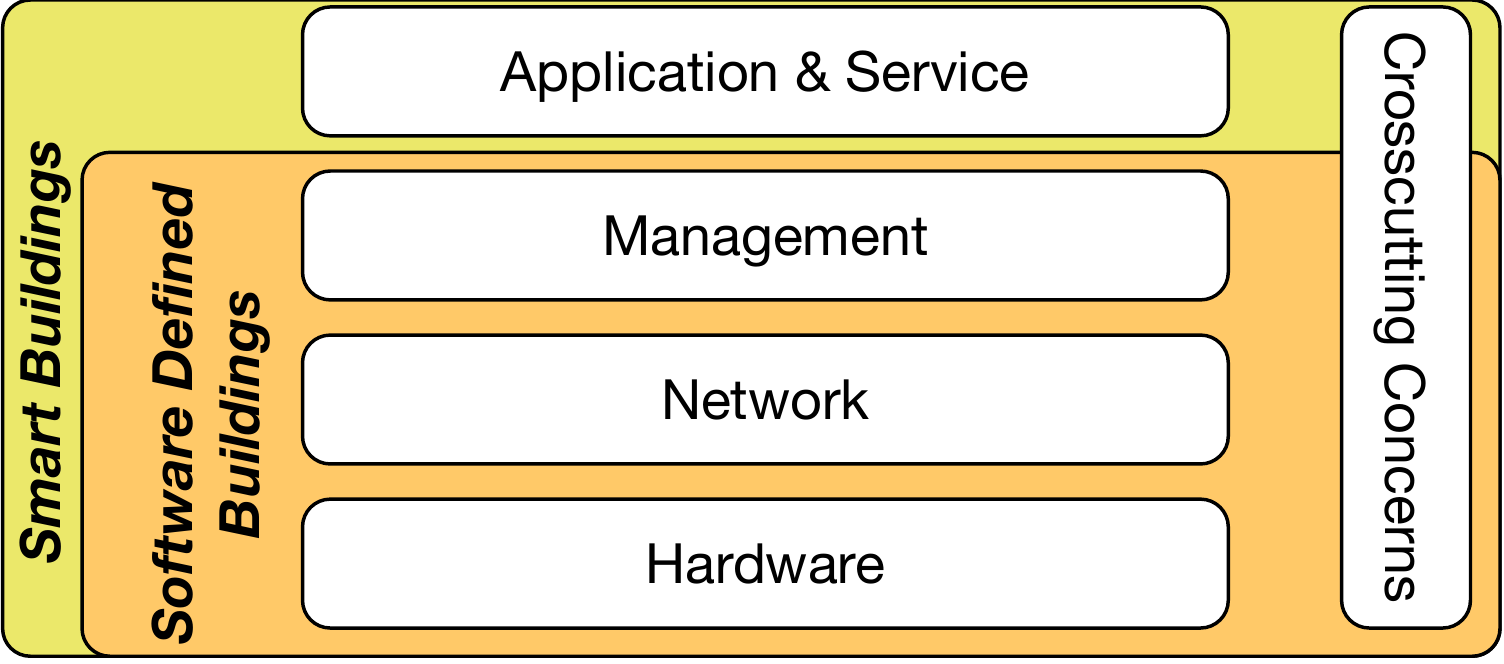}
    \caption{Smart Building ICT Reference architecture}
    \label{fig:arch}
\end{figure}
%From the above considerations, also based on literature in the broad area of smart environments \cite{SCARCH}, IoT \cite{IOTARCH} and smart buildings \cite{SBARCH}, it could be helpful for researchers and practitioners to rationalise all smart buildings concepts and technologies into a reference architecture for ICT smart buildings in the IoT era.

One of the contributions of this paper is the identification of a conceptual frame useful to organize concepts around the idea of SSDB.
We propose a four-layer reference architecture shown in Fig. \ref{fig:arch}. 
At the bottom of the stack, the \textit{Hardware} layer deals with all aspects related to the SSDB ICT infrastructure, ranging from devices, sensors, actuators and  IoT smart objects to network, storage and processing hardware facilities deployed in the building for automation and smart management purposes.
%for smart applications and services. 
Such SSDB nodes, devices, objects and things have to be properly interconnected, exploiting mechanisms and solutions provided by  the \textit{Network} layer.
On top of the latter, the \textit{Management} layer implements facilities for managing Smart Buildings ICT  resources and data, dealing with interoperability issues arising from heterogeneous devices, as well as data management issues.
Such mechanisms, technologies and solutions allow to implement proper applications and services for a smart building such as those related to smart energy, surveillance,  billing, and so on, which are grouped in the \textit{Application \& Service} layer.
Finally, \textit{Crosscutting Concerns} and issues such as security, privacy, AAA, quality of service and similar, spanning the full stack of this Smart Building ICT architecture, are grouped altogether in the corresponding vertical layer.  
All such layers and modules are detailed in following sections.

The reference architecture of Fig. \ref{fig:arch} covers both SDB and SB, highlighting their relationship. 
An SDB can mainly provide to an SB the  ICT infrastructure, thus including the 3 bottom layers and the corresponding crosscutting concerns, while an SB also implements applications and services thus including the corresponding layer.
As stated above, SB can even be deployed on SDB, by mainly injecting specific application codes on the programmable building infrastructure. 
Anyway, the SDB management level is usually more complex than the one of non SDB-powered SB, since it includes specific mechanisms for hardware abstraction through softwarization (API), and control mechanisms on top of this to manage, deploy and orchestrate tasks into pooled resources.
This is often implemented through specific Building Operating Systems.
In the remaining of the paper we will describe in details each layer above introduced.

\section{Hardware}
%Internet of Things}
\label{sec:IoT}

The hardware components that allow the management of the SSDB environment and the data collection are usually referred as \textit{Internet of Things (IoT)}. The advantages of the introduction of this technology seem to overcome the disadvantages and its application becomes easier every year since the price of hardware components is decreasing fast and the power of microprocessors is growing \cite{Salikhov2016b}. %Salikhov2016a
IoT kept growing for several years now \cite{iotanalytics}. 
It has been defined in multiple ways \cite{7473837}, but IoT  is now usually used to refer to \textit{a set of objects that are connected to the internet and can possibly communicate with each other via different protocols.} 
These objects are \textit{sensors, actuators} and, more generally, \textit{embedded systems}.
Latest statistics state that there are around 22 billion  of IoT devices in the the world \cite{connecteddevs} and enable the creation of a vast range of scenarios where devices communicate and cooperate \cite{7370939}. %5579543,
These different scenarios regard many fields of our life, for example home automation, health, transportation and logistics. 
%In this paper we are concerned with the application of IoT for Energy Management for Smart Buildings, we will therefore discuss research and industrial applications emerging in this field.
%\textbf{Ilya, Vivek}
%\section{IoT in Smart Buildings}
%\label{sec:IoTSB}
%\textbf{TODO: Salvatore, Smruti Sarangi}
With regard to smart buildings,  IoT sensing and actuation components can be categorized as follows: 

%\begin{itemize}
%\item 
\textbf{Occupancy Detectors}: these are specialized circuits that are placed in light bulbs, doors, and parking lots. They have dedicated motion sensors for sensing if there are individuals in the vicinity. If they detect motion, then the room is activated, which means switching on light bulbs, opening doors, and marking the parking slot as busy.

%\item 
\textbf{Positioning and Tracking} : these sensors are placed on the person of individuals or are placed at different parts of a room to track the position of a person. They are typically used to track the movements of geriatric patients. We can have specialized {\em activity detectors} that detect if a patient is falling down, or has suddenly become immobile.

%\item 
\textbf{Ambient Control} Building automation is typically used in controlling the environment such as the air conditioning systems, heaters, humidity controllers, and for measuring the pollution, and ambient noise. 
They can be used to optimize the air conditioning cooling/heating only those parts of buildings that are populated. 
Most buildings with centralized air conditioning waste a lot of power because they assume that the entire building is populate at all times.

%\item 
\textbf{Measurement of Usage}: home automation systems are being extensively used as of today to measure the consumption of electricity, gas, and water. The power bill or the water bill based on analyses of usage patterns. 

\textbf{Security}: starting from early burglar alarm systems security has been an important user of home automation technologies.
Modern smart homes have an array of sensors at important entry points, and integrate this information with the motion detected
from CCTV cameras. 
%\end{itemize}

This wide variety of sensors, actuators and devices, is complemented by networking, storage and processing infrastructure facilities such as local switches, routers, NATs, and firewalls for networking, NAS and/or storage servers for storage as well as servers for processing.
Of course, storage and processing servers could even be virtual or remote, provided by Cloud services.

\section{Networking}
\label{sec:net}
%\textbf{TODO: talk about specific network protocols adopted in  Smart Buildings: ZigBee, LoRa, WiFi, Rest, CoAP, AMQP, MQTT, ...}
From a technical point of view, communication between devices is an important aspect in building automation systems \cite{Salikhov2016a}, which can operate at different levels of abstraction. At the same time, a fully distributed system can be implemented in the building, in which the smart nodes interact only with each other to accomplish the task together. Given that the energy for communication is expensive, such peer-to-peer interactions are not preferable. Instead, the standard model is that the IoT nodes send their data to a centralized controller that collates all the data, performs some analysis, and then communicates with other upstream gateways.

Network protocols for Smart Buildings solutions are divided into smart device networks and traditional networks designed for high-speed data transfer. It is reasonable that the protocols of smart home networks will use the protocols already established in wireless sensor networks (WSN) and machine-to-machine communications (M2M).
Since adding advanced features to a protocol increases the cost and reduces usability, developing an attractive protocol is not a trivial task and usually represents a trade-off between cost and performance. In terms of the topology used, the mesh network is the most suitable choice of network topology for wireless communication due to the presence of obstacles in the house, such as walls, furniture, etc. \cite{stojkoska2017}. Double mesh, which means that the network is both wired and wireless, providing a suitable solution for buildings in which a wired home automation system was previously installed. There are many communication networks and protocols designed for exchanging information with multiple devices, components, and sensors. Such communication protocols are created by various organizations, consortia or associations, or can be developed privately when a protocol is applied only by specific manufacturer(s) who must first obtain a license to use it. Commonly used protocols for communication in smart buildings via wireless or wired levels are described below \cite{DEKRA2017, lobaccaro2016}:

%\begin{itemize}
   %\item 
   \textbf{InfraRed Data Association (IrDA)}: simple protocol, usually offering one-way communication. It has a limited range and requires direct visibility of a pair of receiver-transmitter.
   
    %\item 
    \textbf{Ethernet}: Fast and robust wired communication with a range of up to 100m, enabling high noise immunity and the ability to power supply via cable for low power nodes.
    
    %\item 
    \textbf{UWB}: an indoor short-range high-speed wireless communication (up to 10 m) with the bandwidth of over 110 Mbps (up to 480 Mbps), which can satisfy most multimedia applications, such as audio and video delivery in home networks.
        
    %\item 
    \textbf{WiFi}: Fast and reliable wireless IPv6 with a transmission distance of about 25m. Its main feature is the existing broad support: almost every new electronic device comes with WiFi technology installed. As a rule, this is a upper level protocol, where IP is the most predominant protocol that allows communications over the Internet without using a protocol translator.
    
    %\item 
    \textbf{WLAN}: Wireless local area network (WLAN), also known as Wireless Ethernet, is capable to provide reliable communication with low latency for both point-to-point and point-to-multi-point transmissions up to 250m. WLAN applies spread spectrum technology, so users can occupy the same frequency bands with minimal interference to each other.
    
    %\item 
    \textbf{Bluetooth}: a short-range wireless protocol (up to 10 m), the main characteristics of which are low power consumption (especially Bluetooth low energy - BLE) and fast data exchange, as well as widespread availability. Its adaptive frequency hopping system detects existing signals, such as WiFi, and coordinates the channel map for Bluetooth devices to minimize interference. It also implements node discovery services.
 
 % \item 
    \textbf{6LoWPAN}: the IPv6 low power adaptation for devices with limited resources, combining the advantages of both IP and Bluetooth and enabling mesh networks for energy-saving applications in smart buildings with a distance up to 200m.

        %\item 
    \textbf{Thread}: the IPv6-based, low-power technology for IoT networks designed to provide security and meet future requirements. The specification of the Thread protocol is available free of charge, however, this requires consent and permanent adherence to the license agreement. Thread exploits 6LoWPAN, which is based on the use of a connecting router, called an edge router. This means that 6LoWPAN does not know about application protocols and changes that reduces the load on the computing power at the edge routers. Thread is designed to exchange data between devices, even when the WiFi network is turned off.
    
    %\item 
    \textbf{Zigbee}: a wireless mesh network that has proven its efficiency and cost-effectiveness when strengthening and expanding the network, having a transmission distance of 10-75m. ZigBee offers low data rates for personal area networks (PAN). It can be widely used in device control, reliable messaging, building automation, consumer electronics, remote monitoring, healthcare, and many other areas.
    
    %\item
    \textbf{Z-Wave}: a mesh network protocol standard designed for remote control applications in residential and business areas, whose bandwidth is about 6 times lower than for Zigbee. This, however, requires less energy to cover the same range as Zigbee. The main advantages of this technology come from a simple command structure, freedom from domestic interference, a low frequency bandwidth control environment and IP support. Z-Wave has typically 30m indoor range, which extends up to 100 m outdoors.
    
    %\item 
    \textbf{KNX}: one of the most popular open protocols for automation. It operates on several physical levels, such as twisted pair, network power line, infrared, Ethernet and RF channel. Subscribers (devices) connected to the bus (network) can exchange information through a common transmission channel (bus). In this case, the information to be transmitted is packed in a telegram and transmitted via cable from a sensor to actuators. Upon successful transmission and reception, each receiving device confirms the receipt of the telegram, otherwise the transmission is repeated only two more times. If there is no confirmation, the transmission process ends. That is why the KNX protocol is not used in the critical industrial applications. In the decentralized topology, the system does not work from the central unit, which means that each individual unit is connected to the most intelligent device of the KNX ecosystem and does not depend on the functioning of other parts. Therefore, if one unit fails, others may continue to work.
%\end{itemize}

\section{Management}
\label{sec:mgmt}
The Smart Building Management layer is in charge of managing the ICT infrastructure. 
Management functionalities form two groups: resource management and data management.
\subsection{Resource Management}
\label{sec:resmgmt}
Smart buildings are melting pots of different, heterogeneous technologies, especially concerning sensing and actuation devices.
To deal with such heterogeneity issues in Smart buildings, specific solution have been developed, mostly in form of operating systems. 
A first attempt in this direction is implemented by \textit{building management system} (BMS) or \textit{building automation system} (BAS) \cite{BA16} providing an automation solution for controlling heating, ventilation and air conditioning (HVAC), security, fire, power, water supply and elevator systems of a building in a coordinated way.
Extending the scope beyond automation, some  attempts in defining more general \textit{building operating systems} (BOS) have been performed, most of them reported and compared in \cite{xbos}, which also proposes a quite advanced BOS solution named XBOS.
However, among them, one of the most relevant is the Building Operating System Services (BOSS) \cite{boss}, since it defined a new approach to deal with smart buildings: through programmable buildings.
This way, the idea of SDB is slowly affirming in the smart building context, even if its potential is still untapped and should be further investigated and implemented. 
Even the concept of virtualization and virtualized buildings have been recently defined in \cite{SDB2}, but there is still room for \textit{ Building Function Virtualization (BFV)} and new  \textit{Virtual Building Functions (VBF)}.

%En-trakTM Smart Building OS

%\textbf{TODO: talk about resource management issues, i.e. interoperability, pooling scheduling placement orchestration - >  building management system or building automation system (BAS). Building  OS, BOS and BOSS}
 
The basic design philosophy of a building operating system is the plurality of possibly unrelated applications, high requirements of fault tolerance, and a very flexible specification for interacting with a plethora of applications. 
Some of the major issues to address by a BOS are related to interoperability, scheduling, placement, pooling and orchestration.

\subsection{Data Management}
\label{sec:datamgmt}
%\textbf{TODO: talk about    data management issue in SB: Big Data, collection, gathering and storage, filtering fusion preporcessing and processing (edge fog cloud), reasoning (but maybe better in the application logic), ...}

The storage and analysis of Smart Building data is challenging in several ways. First, due to the diversity of systems \cite{Ploennigs} and technologies, the building automation technique faces a long relation with interoperability, leading to data integration concerns \cite{Moreno}. 
Secondly, for a better perception and control of instruments, the density of sensors, promptly increases, generating a vast amount of data.  Bashir et al~\cite{bashir} propose a big data analytic framework for smart buildings. 
Let us divide a typically data processing architecture into several layers.

%\begin{description}
%\begin{itemize}
%\item 
\textbf{Sensor layer}:  This layer consists of sensors that generate data, and record ambient parameters.

%\item 
\textbf{Data storage}:  This information is communicated to routing nodes that collect the data
and store it. Bashir et al. propose a TCP based protocol for communicating data and storing
it in a cloud based database. Often no-SQL databases are used to store such streaming data
such as Apache Flume, and HDFS. 

%\item 
\textbf{Analytics}: Some of the common engines that are used on such platforms are based on
the classic Spark toolkit such as PySpark. 
PySpark can be used to set alerts and thresholds such as: once the level of oxygen dips in a room, oxygen pumps can be automatically started. 

%\item 
\textbf{Rule Engine}: This is an engine that has a set of pre-written rules. Every rule has 
a set of pre-conditions, and a set of actions. If the conditions for firing a rule
exist, then the rule is fired, and appropriate action is taken. 

%\item 
\textbf{Visualization}: The last component, where users can visualize all the elements in the SSDB, the data that  are generated, and the actions that are being performed. 
%\end{itemize}
%\end{description}

\section{Applications, Services and Ambient Intelligence}
%\section{Application and services}
\label{sec:app}

From an end-user/application perspective, the idea of SB belongs to a wider concept: Ambient
Intelligence (AmI) \cite{Sadri11}. 
%This term identifies a vision where the environment supports the people living it by incorporating data acquisition, computation, intelligence and behavior to everyday interconnected objects. 
According to AmI vision, the environment is able to anticipate the needs of its inhabitants therefore responding in a timely and user-friendly way. 
%Technology allows today to recognize users emotion, therefore AmI should also be able to respond accordingly \cite{Fazenda2012}.
Implementing the vision of AmI is highly ambitious, however to reach this objective it is necessary to realize at least a basic set of services that should be part of normal Smart Buildings' behavior. In this section, we will discuss the main application and services that Smart Building should offer towards this vision. We identify three major areas where benefits may soon appear: Energy Management, Environmental Comfort and Safety and Surveillance

%\textbf{TODO: introduce basic services for SB such as reasoning? and others (what?)}

%\textbf{TODO: describe the SB application domain and scenario, from %\cite{buckman2014smart}
%\begin{itemize}
%    \item longevity, maintenance
%    \item energy and efficiency  (apart energy, optimization,  sharing resources and devices, control, ...)
%    \item comfort and satisfaction, well-being, wellness
%    \item  safety,  (I would like to add this that is not present in \cite{buckman2014smart}, but this is one of the main driver of SB, e.g. for risk and emergency management, surveillance, ...)
%\end{itemize}
%}

\subsection{Energy Management}
\label{sec:EM}

%Energy Management is the proactive, organized and systematic management of distribution and energy use in a building to satisfy consumer needs, including both environmental and economic requirements \cite{VDIStandard}. Effective energy management can nowadays be achieved by the use of IoT infrastructure and smart building technologies.
%This paper explores this field and the current solutions, emphasizing the open problems. We will first survey the Internet of Things as a framework capable to offer solutions to multiple application domains, and then we will specifically study the state-of-the-art for what concerns its application to the domain of smart buildings. Subsequently, we shall focus on the specific problems and solutions in energy management, which is only one of the application areas for IoT technology, although particularly important for reasons that we will determine.
Given rising energy costs, energy management in modern buildings is vitally important. Almost all new constructions need be {\em green} buildings. 
The reasons are that modern houses and buildings consume a lot of energy, which can be reduced by efficient management. 
%For example, we can turn off the air conditioners of rooms and offices that are empty, or we can automatically turn off the lights when there is enough ambient light. 
%Such modifications require smart sensors, rule processing engines, and a large number of actuators. 
Consumption in residential buildings accounts for about 40\% of the total energy consumption in the world. 
In comparison, commercial and business buildings (offices, shops, shopping malls, hotels...) and spaces with public functions (schools, hospitals...) use approximately 30\% of energy resources \cite{IoTinnovation}. 
In recent years there has been a growing number of energy regulations and certifications, so the necessity to reduce energy consumption has became more urgent. 
At the same time, rising energy costs have made efficient energy management a matter of survival.

%Smart homes have the technological potential to increase energy efficiency and decrease the costs of energy use \cite{6573315}. 
It is not surprising that in such a growing market the major IT corporations have realized the importance of home and building automation, in particular the aspects that concern energy management. Intel, for example, provides an IoT platform with analytics to offer building operators and managers the possibility of keeping systems functional and cost-effective \cite{IntelEM}. 
The Nokia smart building energy management application has been designed to monitor and control critical building systems and ensure efficient operation.
Reporting and alerts help managers in determining the areas where there is high energy consumption, and the energy use can be optimized \cite{NokiaEM}.
Academic research has also focused on the topic of identifying open issues and possible solutions \cite{ROCHA2015203} and characterizing energy management systems (EMS) \cite{7216662}. An energy management system is defined as \textit{a computer-aided tool used by operators of electric utility grids to monitor, control, and optimize the performance of the generation and/or transmission system}. Some researchers have gone beyond the identification of issues and have proposed approaches and technology stacks, and in fact designed functioning energy management systems for smart homes \cite{SHAKERI2017154}. 
%8246800

A \textit{Home/Building Energy Management System (H/BEMS) is a system that incorporates sensors in household appliances through a home/building network to optimize energy consumption}. 
Most of the academic and industrial solutions are defined according to a general architecture for H/BEMS.
%Following the three-layers organization for home automation presented in Section \ref{sec:SH} and according to current literature, we can determine a general architecture for HEMS. Within the frame defined by this general architecture
%\begin{itemize}
%\item 
This is composed of a \textit{Home/Building Area Network (H/BAN)}, a local residential network interconnecting devices (sensors, smart plugs, intelligent thermostats, cameras). 
It could be a wireless wired network. 
%\item 
Then, \textit{Monitoring and Control Devices} are responsible for monitoring and controlling the energy consumption of appliances and devices.
%\item 
\textit{Processors} are in charge of concentration, storage and management of the information. 
The server and the database are in this central module.
%\item 
\textit{Gateways} ensure connection between the H/BEMS and the outside to allow remote access.
%\end{itemize}

\subsection{Environmental Comfort}
The basic function of Smart Building is to regulate environment parameters such temperature, humidity, lightning  to maximize comfort. This problem has been considered in literature in non-trivial way. For example, in \cite{Park2018} a dynamic thermal model of occupants has been proposed. The model is based on the heat balance equation of human body and thermal characteristics of the occupants. In the context of Smart Offices, occupant satisfaction with indoor environmental conditions has been studied in \cite{Aryal:2018}.
This application area is broader than just regulating temperature and similar parameter. While some of the results of environmental regulation have to be immediate and promptly perceivable, sometime well-being of space inhabitants may depend on factor that require deeper attention. In \cite{Chen:2014} the authors introduce a real system deployed in the offices of four Microsoft campuses in China in order to monitor indoor air quality enabling employees to enquire the air quality of a place by using a mobile phone or checking a website and then make informed decisions.

\subsection{Safety and Surveillance}
Maximal safety conditions are guaranteed by Smart Buildings through surveillance operated by IoT technology. Cameras are just one of the basic instruments of surveillance, but it is certainly not the only one. Access authorization can be performed via biometric parameter and face recognition can be used for security reason. It is important to emphasize that, although the device infrastructure is an important element to ensure safety, it is the software playing a major role here.
In \cite{Amato2018} the authors realized that information provided by single sensor and surveillance technology may not be sufficient to understand the whole context of the monitored environment, and therefore propose the Smart Building Suite, where independent and different technologies are developed in order to realize a multimodal surveillance system. 

\section{Crosscutting concerns}
\label{sec:cc}

%\textbf{TODO: talk about security, performance/timeliness of some (near) real-time app domain app domains (surveillance, safety app, ...), QoS, adaptability (see \cite{buckman2014smart}) ...}

Crosscutting Concerns are issues spanning the full stack of our reference architecture. Each of these would require a separate investigation. Given the limited space we will here scratch the surface and identify two of these concerns:

%\begin{itemize}
%\item 
\textbf{Security}: The more a building is connected the more security threats it is subject to. Given the experiences with automotive or other devices, building owner may be aware of it \cite{DragoniGM16,}. %DonnoDGM16
The problem has to be addressed from two different point of view: technical and communication.

%\item 
\textbf{Adaptability}: Information internally and externally gathered from a range of sources can be used to prepare buildings for a particular event before that event actually happens  \cite{buckman2014smart}. This is a radical shift from the idea of \textit{reactive} building to the one of \textit{proactive}. Smart Buildings should be adaptive. For example a building can adapt to different times of year and different people's perceptions of comfort. 
%\end{itemize}

\section{Conclusions}
\label{sec:conclusions}

In this paper we surveyed SSDB and related technologies, and looked into the major solutions proposed by industry and academia. We would like here to emphasize the major concerns for the widespread adoption of such technologies:

%\begin{itemize}
%\item 
\textbf{Skilled Operators}: people do not know the existence of the described frameworks. In general, operators have no experience and skills in managing Smart Buildings and analyzing large amounts of performance data. Even adopting the technology, it will take a significant training effort before operating buildings in the optimal way \cite{smartop}.

%\item 
\textbf{Costs of startup and maintenance}: installing such technologies and maintaining them is at the moment expensive and we are not aware of tax incentives in most of the countries. Without some kind of incentives it will be hard to convince building owners that the investment is worth the cost, especially for small- and medium-sized buildings \cite{howtoconquer}.

%\item 
\textbf{Interoperability}: as mentioned before in this paper, not all smart devices are interoperable with each other and making things work could be a challenge. Standard protocols to connect all devices do not exist at the moment \cite{8418047}.

%\end{itemize}

There are several aspects that we did not cover here, but that would deserve a separate discussion. First, Autonomous Cars and Smart Buildings are two emerging technologies that are destined to cooperate %Strugar18,
\cite{Strugar19,Gmehlich13}. Second, the use of cryptocurrencies for payments and financial interactions between smart entities will soon emerge as a consuetude, and most likely Blockchain may be the enabling backbone \cite{8191102}.

%\textbf{TODO: Revise and possibly extend}

\bibliographystyle{abbrv}
\bibliography{main}

\begin{thebibliography}{10}

\bibitem{IoTinnovation}
How iot technology is changing building energy management systems.
\newblock
  \url{https://internet-of-things-innovation.com/insights/the-blog/how-iot-technology-is-changing-building-energy-management-systems/}.
\newblock Accessed: 2018-02-10.

\bibitem{howtoconquer}
How to conquer common smart building hurdles.
\newblock
  \url{https://www.buildings.com/article-details/articleid/21098/title/scale-the-smart-technology-barriers}.
\newblock Accessed: 2018-02-10.

\bibitem{connecteddevs}
Internet of things (iot) connected devices installed base worldwide from 2015
  to 2025 (in billions).
\newblock
  \url{https://www.statista.com/statistics/471264/iot-number-of-connected-devices-worldwide/}.
\newblock Accessed: 2018-02-09.

\bibitem{iotanalytics}
Iot platforms: Market report 2015-2021.
\newblock \url{
  https://iot-analytics.com/product/iot-platforms-market-report-2015-2021-3/}.
\newblock Accessed: 2018-02-09.

\bibitem{NokiaEM}
Nokia smart building energy management.
\newblock
  \url{https://internet-of-things-innovation.com/iot-solutions/smart-building-energy-management/}.
\newblock Accessed: 2018-02-09.

\bibitem{ElectronicAutomotive}
The role of electronics in automotive advancements.
\newblock
  \url{https://auto.howstuffworks.com/car-driving-safety/safety-regulatory-devices/new-technologies-cars-less-safe1.html}.
\newblock Accessed: 2018-02-09.

\bibitem{IntelEM}
Smart building energy management solutions.
\newblock
  \url{https://www.intel.com/content/www/us/en/energy/solutions/smart-building-energy-management.html}.
\newblock Accessed: 2018-02-09.

\bibitem{SMForbes}
Smart buildings: Forming the foundation of smart cities.
\newblock
  \url{www.forbes.com/sites/insights-inteliot/2018/10/24/smart-buildings-forming-the-foundation-of-smart-cities/}.
\newblock Accessed: 2018-02-09.

\bibitem{DEKRA2017}
Smart home protocols explained.
\newblock
  \url{https://medium.com/iotforall/smart-home-protocols-thread-zigbee-z-wave-knx-and-more-71efa4b410e1}.
\newblock Accessed: 2018-02-09.

\bibitem{smartop}
Smart operators for smart buildings.
\newblock
  \url{http://www.cunybpl.org/wp-content/uploads/2017/11/Smart-Operators-Smart-Buildings.pdf}.
\newblock Accessed: 2018-02-10.

\bibitem{SDB2}
M.~A. {Abid} and H.~{De Meer}.
\newblock Virtualized software defined buildings: a key enabler of the future
  smart cities.
\newblock In {\em SmartGridComm}, 2018.

\bibitem{Amato2018}
G.~Amato, P.~Barsocchi, F.~Falchi, E.~Ferro, C.~Gennaro, G.~R. Leone,
  D.~Moroni, O.~Salvetti, and C.~Vairo.
\newblock Towards multimodal surveillance for smart building security.
\newblock {\em Proceedings}, 2(2), 2018.

\bibitem{Aryal:2018}
A.~Aryal, F.~Anselmo, and B.~Becerik-Gerber.
\newblock Smart iot desk for personalizing indoor environmental conditions.
\newblock In {\em 8th Int. Conf. on IoT}, IOT '18, pages 35:1--35:6. ACM, 2018.

\bibitem{bashir}
M.~R. Bashir and A.~Q. Gill.
\newblock Towards an iot big data analytics framework: smart buildings systems.
\newblock In {\em Int. Conf. on HPCC/SmartCity/DSS}, pages 1325--1332. IEEE,
  2016.

\bibitem{buckman2014smart}
A.~Buckman, M.~Mayfield, and S.~BM~Beck.
\newblock What is a smart building?
\newblock {\em Smart and Sustainable Built Environment}, 3(2):92--109, 2014.

\bibitem{Chen:2014}
X.~Chen, Y.~Zheng, Y.~Chen, Q.~Jin, W.~Sun, E.~Chang, and W.-Y. Ma.
\newblock Indoor air quality monitoring system for smart buildings.
\newblock In {\em Int. Conf. on Perv. \& Ubiq. Comp.}, UbiComp, pages 471--475.
  ACM, 2014.

\bibitem{boss}
S.~Dawson-Haggerty, A.~Krioukov, J.~Taneja, S.~Karandikar, G.~Fierro,
  N.~Kitaev, and D.~Culler.
\newblock $\{$BOSS$\}$: Building operating system services.
\newblock In {\em USENIX Net. Sys. Design \& Impl. (NSDI 13)}, pages 443--457,
  2013.

\bibitem{SDB1}
S.~{Dawson-Haggerty}, J.~{Ortiz}, J.~{Trager}, D.~{Culler}, and R.~H. {Katz}.
\newblock Energy savings and the “software-defined” building.
\newblock {\em IEEE Design Test of Computers}, 29(4):56--57, Aug 2012.

\bibitem{BA16}
P.~Domingues, P.~Carreira, R.~Vieira, and W.~Kastner.
\newblock Building automation systems: Concepts and technology review.
\newblock {\em Computer Standards \& Interfaces}, 45:1--12, 2016.

\bibitem{DragoniGM16}
N.~Dragoni, A.~Giaretta, and M.~Mazzara.
\newblock The internet of hackable things.
\newblock In {\em {SEDA}}, volume 717 of {\em Adv. in Intel. Sys. \& Comp.},
  pages 129--140. Springer, 2016.

\bibitem{xbos}
G.~Fierro and D.~E. Culler.
\newblock Xbos: An extensible building operating system.
\newblock In {\em BuildSys}, pages 119--120. ACM, 2015.

\bibitem{Gmehlich13}
R.~Gmehlich, K.~Grau, A.~Iliasov, M.~Jackson, F.~Loesch, and M.~Mazzara.
\newblock Towards a formalism-based toolkit for automotive applications.
\newblock In {\em Formal Methods in Software Engineering (FormaliSE)}, 2013.

\bibitem{8418047}
A.~Khusnutdinov, D.~Usachev, M.~Mazzara, A.~Khan, and I.~Panchenko.
\newblock Open source platform digital personal assistant.
\newblock In {\em WAINA}, 2018.

\bibitem{8191102}
C.~Lazaroiu and M.~Roscia.
\newblock Smart district through iot and blockchain.
\newblock In {\em ICRERA}, pages 454--461, Nov 2017.

\bibitem{lobaccaro2016}
G.~Lobaccaro, S.~Carlucci, and E.~L{\"o}fstr{\"o}m.
\newblock A review of systems and technologies for smart homes and smart grids.
\newblock {\em Energies}, 9(5):348, 2016.

\bibitem{7370939}
P.~Mekikis, A.~Antonopoulos, E.~Kartsakli, A.~S. Lalos, L.~Alonso, and
  C.~Verikoukis.
\newblock Information exchange in randomly deployed dense wsns with wireless
  energy harvesting capabilities.
\newblock {\em IEEE Transactions on Wireless Communications}, 15(4):3008--3018,
  April 2016.

\bibitem{SDC}
G.~{Merlino}, D.~{Bruneo}, F.~{Longo}, A.~{Puliafito}, and S.~{Distefano}.
\newblock Software defined cities: A novel paradigm for smart cities through
  iot clouds.
\newblock In {\em 2015 IEEE 12th Intl Conf on Ubiquitous Intelligence and
  Computing and 2015 IEEE 12th Intl Conf on Autonomic and Trusted Computing and
  2015 IEEE 15th Intl Conf on Scalable Computing and Communications and Its
  Associated Workshops (UIC-ATC-ScalCom)}, pages 909--916, Aug 2015.

\bibitem{Moreno}
M.~Moreno.
\newblock Applicability of big data techniques to smart cities deployments.
\newblock {\em IEEE Trans. on Industrial Informatics}, 13:800--809, 2017.

\bibitem{Nalin2016}
M.~Nalin, I.~Baroni, and M.~Mazzara.
\newblock A holistic infrastructure to support elderlies' independent living.
\newblock {\em Encyclopedia of E-Health and Telemedicine, IGI Global}, 2016.

\bibitem{Park2018}
H.~Park and S.-B. Rhee.
\newblock Iot-based smart building environment service for occupants’ thermal
  comfort.
\newblock {\em Journal of Sensors}, 2018.

\bibitem{Ploennigs}
J.~Ploennigs.
\newblock A modular, adaptive building automation system ontology.
\newblock In {\em 38th Ann. Conf. on IEEE Industrial Electronics Society}.
  IEEE, 2012.

\bibitem{ROCHA2015203}
P.~Rocha, A.~Siddiqui, and M.~Stadler.
\newblock Improving energy efficiency via smart building energy management
  systems: A comparison with policy measures.
\newblock {\em Energy and Buildings}, 88:203 -- 213, 2015.

\bibitem{Sadri11}
F.~Sadri.
\newblock Ambient intelligence: {A} survey.
\newblock {\em {ACM} Comput. Surv.}, 43(4):36:1--36:66, 2011.

\bibitem{Salikhov2016b}
D.~Salikhov, K.~Khanda, K.~Gusmanov, M.~Mazzara, and N.~Mavridis.
\newblock Jolie good buildings: Internet of things for smart building
  infrastructure supporting concurrent apps utilizing distributed
  microservices.
\newblock In {\em 1st Int. Conf. "Convergent"}, pages 48--53, 2016.

\bibitem{Salikhov2016a}
D.~Salikhov, K.~Khanda, K.~Gusmanov, M.~Mazzara, and N.~Mavridis.
\newblock Microservice-based iot for smart buildings.
\newblock In {\em Int. Conf. WAINA}, 2017.

\bibitem{7473837}
J.~Santos, J.~J. P.~C. Rodrigues, J.~Casal, K.~Saleem, and V.~Denisov.
\newblock Intelligent personal assistants based on internet of things
  approaches.
\newblock {\em IEEE Systems Journal}, 12(2):1793--1802, June 2018.

\bibitem{7216662}
I.~D. Serna-Suárez, G.~Ordóñez-Plata, and G.~Carrillo-Caicedo.
\newblock Microgrid's energy management systems: A survey.
\newblock In {\em EEM}, pages 1--6, 2015.

\bibitem{SHAKERI2017154}
M.~Shakeri, M.~Shayestegan, H.~Abunima, S.~S. Reza, M.~Akhtaruzzaman,
  A.~Alamoud, K.~Sopian, and N.~Amin.
\newblock An intelligent system architecture in home energy management systems
  (hems) for efficient demand response in smart grid.
\newblock {\em Energy\&Buildings}, 138:154--164, 2017.

\bibitem{stojkoska2017}
B.~L.~R. Stojkoska and K.~V. Trivodaliev.
\newblock A review of internet of things for smart home: Challenges and
  solutions.
\newblock {\em Journal of Cleaner Production}, 140:1454--1464, 2017.

\bibitem{Strugar19}
D.~Strugar, R.~Hussain, M.~Mazzara, V.~Rivera, I.~Afanasyev, and J.~Lee.
\newblock An architecture for distributed ledger-based m2m auditing for
  electric autonomous vehicles.
\newblock In {\em {33rd Int. Conf. WAINA}}, 2019.

\end{thebibliography}

\end{document}